\documentclass{sigkdd}

\usepackage{amsmath,amssymb}
\usepackage{csquotes}
\usepackage{wrapfig}
\usepackage{multirow}
\usepackage{url}
\usepackage{algorithm}
\usepackage{algpseudocode}
\usepackage{hyphenat}
\usepackage{graphicx}
\usepackage{lipsum} 
\usepackage{subcaption} 
\usepackage{adjustbox} 
\usepackage[svgnames]{xcolor} 
\usepackage{pgfplots} 
\usepackage{tabularx}
\usepackage{booktabs}
\usepackage{bm}
\usepackage{arydshln}
\usepackage{mathrsfs}
\usepackage{afterpage}
\usepackage{float} 
\usepackage[normalem]{ulem} 
\usepackage{ulem}
\usepackage{stackengine}
\usepackage{placeins}
\usepackage{pgfplots}
\usepgfplotslibrary{groupplots}
\usepackage{titlesec}
\usepackage{placeins}
\usepackage{stfloats}
\usepackage{pgfplots}
\pgfplotsset{compat=1.18}  
\titleformat{\section}{\normalfont\Large\bfseries}{\thesection}{1em}{}

\titlespacing{\section}{0pt}{*0}{*0} 

\begin{document}

\title{Is Less Really More?\\Fake News Detection with Limited Information}
%

\numberofauthors{3}
%


\author{
%
\alignauthor Zhaoyang Cao \\
       \affaddr{Data Lab, EECS Department}\\
      \affaddr{Syracuse University}\\
       \email{zycao@data.syr.edu}
\alignauthor John Nguyen\\
       \affaddr{Syracuse University}\\
       \email{jnguye30@syr.edu}
\alignauthor Reza Zafarani\\
     \affaddr{Data Lab, EECS Department}\\
      \affaddr{Syracuse University}\\
       \email{reza@data.syr.edu}
}
\date{}

\maketitle
\begin{abstract}
The threat that online fake news and misinformation pose to democracy, justice, public confidence, and especially to vulnerable populations has led to a sharp increase in the need for fake news detection and intervention. Whether multi-modal or pure text-based, most existing fake news detection methods depend on textual analysis of entire articles. However, these fake news detection methods come with certain limitations. For instance, fake news detection methods that rely on full text can be computationally inefficient, demand large amounts of training data to achieve competitive accuracy, and may lack robustness across different datasets. This is because fake news datasets have strong variations in terms of the level and types of information they provide; where some can include large paragraphs of text with images and metadata, and others can be a few short sentences. Perhaps if one could only use minimal information to detect fake news, fake news detection methods could become more robust and resilient to the lack of information. We aim to overcome these limitations by detecting fake news using systematically selected, limited information that is both effective and capable of delivering robust, promising performance. We propose a framework called \textsf{SLIM} (\textbf{S}ystematically-selected \textbf{L}imited \textbf{I}nfor\textbf{m}ation) for fake news detection. In \textsf{SLIM}, we quantify the amount of information by introducing information-theoretic measures. \textsf{SLIM} leverages limited information (e.g., a few named entities) to achieve performance in fake news detection comparable to that of state-of-the-art obtained using the full text, even when the dataset is sparse. Furthermore, by combining various types of limited information, \textsf{SLIM} can perform even better while significantly reducing the quantity of information required for training compared to state-of-the-art language model-based fake news detection techniques.
\end{abstract}

\newcommand{\zy}[1]{\textcolor{red}{#1}}
\section{Introduction}\flushbottom
The demand for fake news detection and intervention has grown rapidly due to the threat that false news poses to democracy, justice, and public confidence~\cite{harriss2017online,shu2017fake,zhou2020survey}. Among several fake news detection methodologies, research has shown that advanced pre-trained large language models and multimodal frameworks perform significantly better than traditional machine learning and deep learning models. Language models perform better as they can learn contextual text representations during pretraining \cite{khan2021benchmark}. One example is the work by Bhatt \textit{et al.}, which proposes a Siamese network framework with multiple branches built on the BERT architecture, where each branch is tailored to process distinct types of textual information (such as article bodies, and social media comments). Using this enhanced sequence model, the framework can achieve a competitive performance on fake news detection~\cite{bhatt2022fake}. It has also been shown that systems that combine topical distributions (e.g., from Latent Dirichlet Allocation) with text representations from large language models perform well on fake news detection \cite{gautam2021fake}.
A multi-modal example is \textsf{SAFE}, which identifies fake news using textual and visual modalities. \textsf{SAFE} analyzes the semantic and visual consistency between news articles (text) and their accompanying images. Harnessing multi-modal information, SAFE enhances the accuracy of fake news detection across different media formats \cite{zhou2020similarity}. These fake news detection techniques mostly rely on the textual analysis of the entire text as the primary signal for fake news identification, whether they are pure text-based or multi-modal. 

Despite the significant successes of the aforementioned methods, we still cannot neglect some drawbacks of the full text-based approaches. One of the primary concerns is computational efficiency. In addition, full text may not always be available in datasets used to train fake news detection models. The negative impact on efficiency is particularly important in various application scenarios, especially those requiring real-time responses. Consequently, \textit{relying on limited information to detect fake news is a more competitive option in practical applications}, as it significantly decreases computational complexity while remaining robust and efficient to data sparsity constraints~\cite{raza2022fake,shu2017fake}. 

However, merely reducing the amount of information is not sufficient, but such limited information should be strategically identified to maintain effectiveness, and one cannot simply rely on better machine learning techniques or better large language models. In particular, while large language models have achieved promising results on fake news detection, future language models, as noted by Tamkin \textit{et al.}, might make it difficult or impossible to identify disinformation when only relying on the text body of the news article~\cite{tamkin2021understanding}. Research has shown that humans can be deceived by news produced by the GPT-2 and other language models and human detection is expected to become more challenging \cite{solaiman2019release}. ``Full-text''-based detection techniques would be insufficient as advanced language models mimic the real distribution of human text~\cite{solaiman2019release}. Hence, while language models have been widely proven to outperform other generic models in fake news detection, we cannot neglect that the rapid growth of language models will hinder human detection. As a result, the difficulty of identifying disinformation motivates research to rely more heavily on other limited yet subtle information cues in fake news articles. Such subtle cues will play an essential role in detecting online malicious activity, as also noticed by other research studies \cite{solaiman2019release, tamkin2021understanding}. But what are examples of such limited information cues? By surveying the literature~\cite{arora2022metadata,solaiman2019release, tamkin2021understanding,zhou2020survey}, we categorize such limited cues into three broad types: (a) \textit{keywords}, (b) \textit{sequences}, and (c) \textit{metadata}.

Researchers have explored improving fake news detection by harnessing such limited information cues~\cite{arora2022metadata,vincent2023personalised,zhang2021match}. However, these efforts face two key challenges: (1) the approaches primarily \textit{integrate} these cues (as extra machine learning features) with existing “full-text”-based models, making it unclear how limited information alone contributes to fake news detection; and (2) the integrations are often ad-hoc and rely heavily on feature engineering, leaving open questions about which types (or quantities) of limited information are most beneficial. Our goal in this paper is to address these challenges.\vspace{1mm}

\noindent \textbf{This paper: Fake Detection with Limited Information.} We aim to identify the means to utilize limited information for fake news detection through a systematic analysis of various ways of extracting information from limited information (e.g., keyword extraction and sequence tagging). To ensure that, in fact, less information is used, we propose information-theoretic measures to assess information quantity. Subsequently, we explore how various types of key limited information can be combined. We utilize this newly identified key information as input in a language model to assess its impact on the effectiveness of fake news detection and broadly investigate the following research direction (details can be found in Section \ref{expe}): 1. We assess the impact of different types of limited information on fake news detection; 2. We study the influence of multiple modalities of limited information on fake news detection; and 3. We compare the performance of utilizing limited information state-of-the-art models. In sum, our major contributions are:
\begin{itemize}
\item[$\blacktriangleright$] To the best of our knowledge, this work is the first to propose various quantified strategies for using limited information for fake news detection.
\item[$\blacktriangleright$] We identified the optimal combinations of utilizing limited information yielding the highest detection accuracy by integrating various key pieces of information. Examples include combining keywords with sequence tagging or keywords with metadata.
\item[$\blacktriangleright$]  We explored the viability of using limited data as a substitute for text body in the realm of fake news detection. All codes are publicly available.\footnote{The code and data is available at \mbox{\url{https://github.com/kappakant/SLIM}}}
\end{itemize}

Section 2 formally presents the related work. Section 3 describes the proposed architecture of the \textsf{SLIM} framework, followed by framework evaluation and experiments that address our research questions in Section 4. Section 5 concludes this research with directions for future work.

\section{Related Work}
We categorize limited information into three main types: keywords, sequences (e.g., POS, NER annotations), and metadata (e.g., titles, authors). This categorization is both (\ref{2.1}) theoretically grounded and (\ref{2.2}) empirically validated, as we will present next. 
\subsection{\textbf{Theoretical Justification}}\label{2.1}
This systematic selection is supported by extensive research in computational linguistics and information retrieval, demonstrating that these information sources provide a comprehensive representation of textual data for downstream tasks (e.g., fake news detection). 
\newline First, the use of keywords is well-supported in computational linguistics and information retrieval for fake news detection. Keywords capture salient lexical features that are often indicative of deceptive or manipulative texts. For instance, Pérez-Rosas \textit{et al.}, showed in their experiments that certain keyword patterns, including sensational phrases or exaggerated emotional expressions, are powerful indicators of fake news, with high classification accuracy \cite{perez2017automatic}. Similarly, keyword-based retrieval, such as those described by Manning \textit{et al.}, \cite{schutze2008introduction}, has been foundational in identifying misinformation documents. 
\newline Sequence tags provide syntactic and semantic structure to text, which is useful for detecting inconsistencies in fake news. Sousa-Silva highlighted that fake news often contains anomalous syntactic patterns, such as inconsistent verb tenses, which can be effectively captured by POS tagging \cite{sousa2022fighting}. NER helps identify entities that are frequently manipulated or misrepresented in fake news \cite{shu2017fake}.
\newline Finally, metadata plays a critical role in assessing credibility. Titles summarize the primary claim of a news article, and their linguistic features, such as clickbait patterns, have been studied by Kong \textit{et al.}, in the context of fake news detection \cite{kong2020fake}. Author has been used by Castillo \textit{et al.}, in their paper, demonstrating its importance in distinguishing reliable sources \cite{castillo2011information}. Together, these information sources—keywords, sequence tags, and metadata—form a comprehensive and robust foundation for fake news detection.
\subsection{\textbf{Empirical Justification}}\label{2.2} These three types of information have also been empirically validated, demonstrating their critical role in downstream tasks, such as fake news detection. In addition, these types can be combined in various capacities to form other types of limited information. We first review the related work on each type of information.
\subsubsection{Keywords}
Keywords are words that precisely and simply characterize an aspect of a subject stated in a document. They are crucial indicators of important textual information that spread among individuals~\cite{siddiqi2015keyword}. Keywords can be extracted from textual documents using a variety of techniques, including statistical, rule-based, machine learning, or domain-specific approaches \cite{siddiqi2015keyword, bharti2017automatic}. However, to ensure that the extracted keywords are semantically consistent with the document, language model-based approaches that handle text to extract keywords can consider contextual information. As a result, the language models' generated keywords might more accurately represent the content of the original text~\cite{grootendorst2020keybert}.

While keywords have been commonly used in fake news detection, systematic research on ways or how to use keywords is relatively lacking. Souza \textit{et al.} proposed the Positive and Unlabeled Learning with the network-based Label Propagation (PU-LP) algorithm, which incorporates a keywords attention mechanism~\cite{de2024keywords}. They employed Yake to extract keywords and then used these keywords in Graph Attention Neural Event Embedding (GNEE) to classify unlabeled nodes. Additionally, due to the unstructured texts of news on certain social media platforms, such as Twitter, Jayasiriwardene and Ganegoda utilized Core NLP and TF-IDF to extract keywords for more effective data collection for fake news detection. Additionally, to improve the precision and effectiveness of relevant news retrieval, they also used the WordNet lexical database to find synonyms and bigrams to generate proper key phrases \cite{jayasiriwardene2020keyword}. 

\subsubsection{Sequences}
\textit{Sequence tagging}, a fundamental task in natural language processing (NLP), involves the assignment of labels to individual tokens in a given sequence, such as words or subwords. These labels typically represent linguistic properties or semantic categories, facilitating various NLP tasks, including Part-Of-Speech tagging (POS), Named Entity Recognition (NER), and chunking. The significance of sequence tagging lies in its ability to discern syntactic roles, semantic entities, and even higher-order linguistic features by analyzing the sequential context of tokens. Furthermore, sequence tagging has great potential for detecting fake news. By leveraging its capacity to identify named entities and recognize linguistic patterns, sequence tagging can assist in the identification of fake information and misleading content~\cite{kapusta2021using, spalenza2020using}. 

\textbf{POS tagging:}
Some researchers have attempted to leverage sequence tagging methods for fake news detection. For instance, Balwant proposed an architecture that combines POS tag information from news articles using bidirectional long short-term memory (LSTM) and author profile information by convolutional neural network (CNN)~\cite{balwant2019bidirectional}. His hybrid architecture showed high performance on the \texttt{LIAR} dataset. According to \cite{pak2010twitter}, certain POS tags are powerful indicators of emotional texts. For example, comparative adjectives (JJR) typically provide information or state facts, whereas superlative adjectives (JJS) are frequently used to express opinions. Positive text commonly features superlative adverbs (RBS) such as ``most'' and ``best.'' In addition, the choice of adjectives and adverbs can alter the meaning and semantics of a sentence. Pairing the same noun or verb with different adjectives or adverbs may result in different interpretations. However, such systematic combinations of POS tags in addition to how much and how often they are helpful have less been explored in research.  The \textsf{SLIM} framework studied in this research will target such research gaps.

\textbf{NER tagging:} NER tags are also used for fake news detection. For instance, Al-Ash and Wibowo improved the BERT model by joining a NER and relational features classification (RFC) into a single formulation \cite{al2018fake}. To improve generalization performance in joint learning, RFC and NER models shared the parameter layer in the BERT-joint framework. Shishah has introduced an approach to vector representation, which incorporates term frequency, inverse document frequency, and NERs~\cite{shishah2021fake}. However, the final results demonstrate that only term frequency yields the best performance when using an SVM classifier. This outcome may be due to the absence of more advanced classifiers or the lack of a proper understanding of crucial information that might be useful in specific NERs. To address such issues, \textsf{SLIM} utilizes language models to extract varying percentages of keyword information and integrates them with proper sequence tags to detect fake news.

\subsubsection{Metadata}
Metadata is often used in fake news detection, where the common approach is to combine it (as extra features) with the full-text body and use it as input for fake news detection in \textit{content-based fake news detection}~\cite{antoun2020state, krevsvnakova2019deep}. Content-based methods are often considered as the traditional approach to detect fake news, an area where researchers have made significant contributions~\cite{guacho2018semi,pan2018content,zhou2020fake,zhou2020survey}. For instance, Wynne and Wint showed that highly accurate fake news classifiers can be trained using Gradient Boosting Classifiers and character $n$-grams as features in experiments~\cite{wynne2019content}. Zhou \textit{et al.} introduced the \textsc{SAFE} model, which investigates the multi-modal content (comprising textual and visual information) of news articles. Their case studies validate the effectiveness of the cross-modal relationship between both textual and visual features of news content  \cite{zhou2020similarity}. 

A few studies have explored the role of metadata in fake news detection. For example, Elhadad \textit{et al.} presented a novel approach to processing the entire textual content of news by extracting various textual features and a complex set of additional metadata-related features without dividing the news documents into sections~\cite{elhadad2020novel}. They employ TF-IDF in the feature extraction phase. Similarly, Amine \textit{et al.} utilized word embedding techniques and convolutional neural networks for feature extraction and compared various deep learning architectures applied to different metadata \cite{amine2019merging}. It is worth noting that past research did not consider the independent impact of metadata and it was always used as an add-on to improve fake news detection. Furthermore, metadata was often preprocessed using vectorizations such as TF-IDF or deep learning; hence, despite being a crucial and valuable limited piece of information, metadata is frequently underexplored.

Differing from existing works, \textsf{SLIM} explores various aspects of metadata, such as whether metadata can replace text and whether it can be augmented by other types of limited information, such as keywords (or sequence-tagging words), for detecting fake news.

\subsubsection{Combining Various Types of Limited Information}
Few studies have integrated the various types of limited information to tackle fake news detection. In a recent paper, Migyeong Yang \textit{et al.} proposed a deep learning approach to debunk fake news about COVID-19 at its early stages  \cite{yang2024fighting}. They designed three embedding layers, the second of which is the Propagated Information Encoder (PIE). In this layer, they used NER tagging words and keywords to extract information for searching related YouTube videos. The text information from these videos, such as titles and descriptions, was then refined and used as input for this layer.

Although their experiment, as a case study, successfully detected fake news on newly emerging and critical topics, it did not provide insights into where and how much limited information is necessary for fake news detection. Furthermore, they only used NER tagging words and keywords as a basis for searching videos rather than integrating these elements as the final set of features for detection.

\section{The \textsf{SLIM} Framework}\label{methodology}

In the following subsections, we will first introduce the problem statement and framework formulation. Next, we will explain the approach to integrating information and performing the downstream task of fake news detection. 

\subsection{Problem Statement and Framework  Formulation}\label{formulation}

Given an ordered set of the news article $A = \{w_1,w_2,...,w_p\}$, where $w_i$ is the $i$th word, $p$ is the total number of words in the article $A$. Our goal is to predict whether $A$ is a fake news article ($\hat{y} = 0$) or a
true one ($\hat{y} = 1$) by investigating its systematically-selected limited information.
\newline\newline$\textsf{SLIM}$ \textbf{Variations}: We have four variations of $\textsf{SLIM}$ based on the types of inputs that each variation takes. Variations of the framework represent the different systematically selected features of information. For notation clarity, we define them as $\textsf{SLIM}_{\textsc{keyword}}$, $\textsf{SLIM}_{\textsc{sequence}}$, $\textsf{SLIM}_{\textsc{metadata}}$, and $\textsf{SLIM}_{\textsc{multimodal}}$. In the following sections, we will introduce the preprocessing steps required to build these variations.

\subsubsection{$\textsf{SLIM}_{\textsc{keyword}}$}\label{kwextract}
\begin{figure}[t]
  \centering  \includegraphics[width=\linewidth]{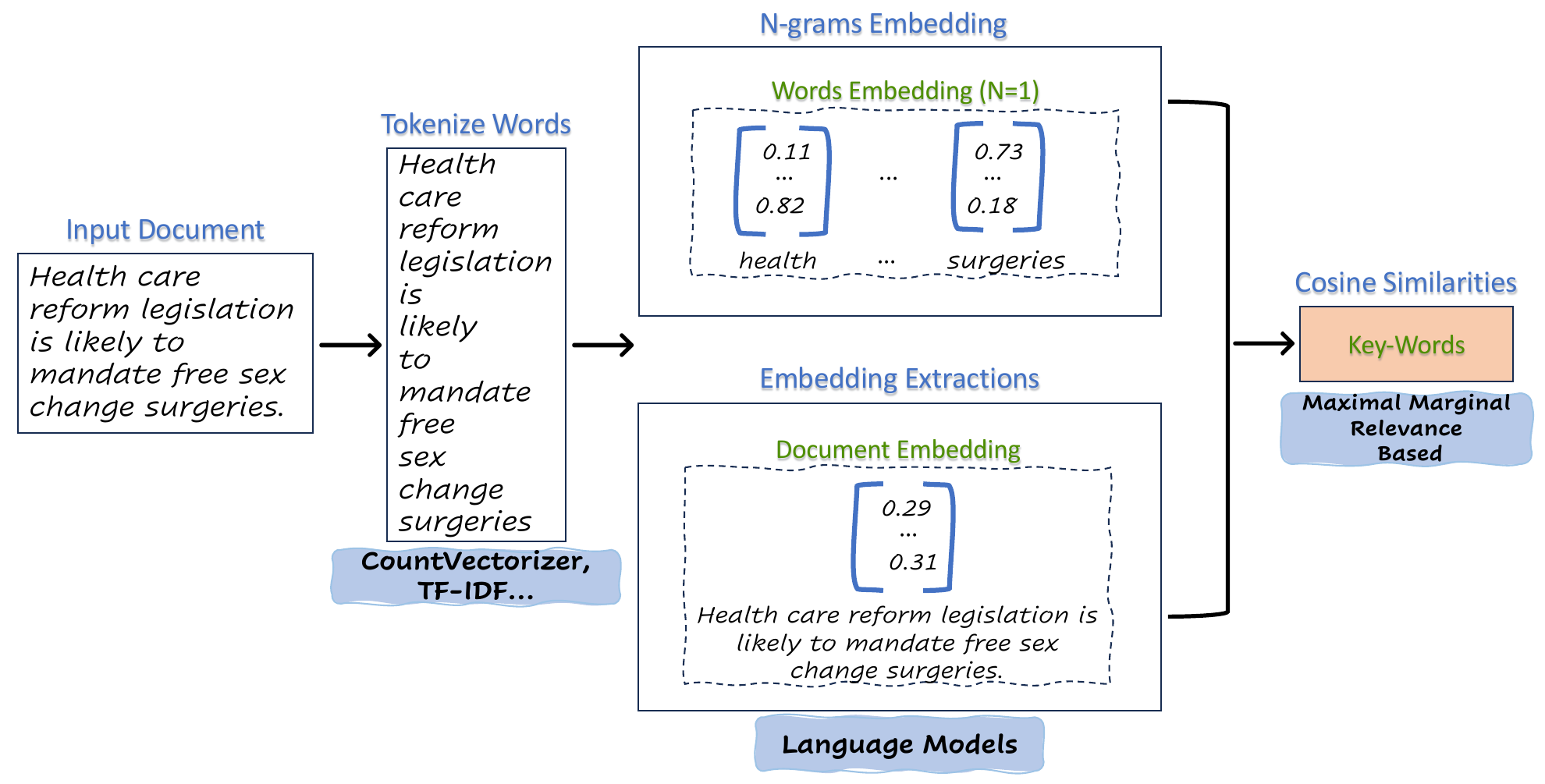}
  \caption{Extracting Keyword Information: the input is the body of the news under the proposed \textsf{SLIM} framework}
  \label{keywords_frame} \vspace{-2mm}
\end{figure}

The first variation $\textsf{SLIM}_{\textsc{keyword}}$ takes keywords as input. The process of extracting keyword information is depicted in Figure \ref{keywords_frame}. 
It includes varying percentages of keywords. To obtain the $\textsf{SLIM}_{\textsc{keyword}}$, we first use BERT to obtain the document embedding  $e_d \in \mathbb{R}^{n}$. Meanwhile, we use the N-grams for word embeddings. When $N=1$, we can get word embedding $e_{w_i}$ for an arbitrary $i_{th}$ word. Then we calculate the cosine similarity (denoted as $S_{cosine}$, given in Equation \ref{equcos}) between document embedding $e_d$ and each $e_{w_i}$ and retain the set of words with a cosine similarity greater than 0. We constrain the extraction process by the maximal marginal relevance (MMR) to avoid the redundancy of the sorting results and to ensure the correlation of the words (stated in Equation \ref{eqummr}). The process of MMR is summarized in Algorithm 1. Formally, the final input $\textsf{SLIM}_{\textsc{keyword}}$ is the set of keywords defined by \begin{equation}\label{equ1}
\textsf{SLIM}_{\textsc{keyword}} =  \{w_i | S_{cosine} (e_{w_i},e_d) > 0\}
,\end{equation} \par
\begin{equation}\label{equcos}
S_{cosine}
(\mathbf{e_d}, \mathbf{e_{w_i}}) = 
\frac{\mathbf{e_d} \cdot \mathbf{e_{w_i}}}{\|\mathbf{e_d}\| \cdot \|\mathbf{e_{w_i}}\|}
,\end{equation}
\par
\begin{equation}\label{eqummr}
\resizebox{\linewidth}{!}{$
\textit{MMR}(e_d,C,R) = \arg\max_{w_i \in C} \Big[ 
    \lambda sim(e_d,e_{w_i}) - (1-\lambda) \max_{w_j \in R} sim(e_{w_i},e_{w_j}) \Big]
$}
\end{equation}

where $e_d$ is the document embedding, $C$ is the set of collected words, $R$ is the returned result set, $e_{w}$ is the word embedding, and $sim$ refers to the cosine similarity $S_{cosine}$. At last, $\lambda$ is the diversity and set to 0.5. Finally, $k$ (in Algorithm 1) is the proportion of the desired number of words relative to the total number of words in the full text. By adjusting the value of $k$, we can derive keywords with the desired varying word counts.

\begin{algorithm}[t]\label{algo1}
  \caption{MMR in $\textsf{SLIM}_{\textsc{keyword}}$}\label{euclid}
  \begin{algorithmic}[1]
    \State Input: C, $|A|$, $e_d$, $e_{w}$, k
    \State Output: R
      \State Initialization: R = $\emptyset$, C = $\textsf{SLIM}_{\textsc{keyword}}$ (set of words that satisfy Equation \ref{equ1})
      \While{$|R| < \lfloor |A| \cdot k \rfloor$}
       \State \resizebox{\linewidth}{!}{$w^*=\arg\max\limits_{w_i \in C}\Big[ 
    \lambda sim(e_d,e_{w_i})-(1-\lambda) \max\limits_{w_j} sim(e_{w_i},e_{w_j}) \Big]$}
        \State $R \leftarrow R \cup \{w^*\}, C \leftarrow C \backslash \{w^*\} $
      \EndWhile\label{euclidendwhile}
      \State \Return R\
  \end{algorithmic}
\end{algorithm}


\subsubsection{$\textsf{SLIM}_{\textsc{sequence}}$}   
In $\textsf{SLIM}_{\textsc{sequence}}$, the framework uses both POS and NER tags as input; the input comprises sets of words from different sequence taggings. For POS tagging, we initially tokenize the news articles. Once we obtain the corresponding tokens, we employ the \textit{pos\_tag} function for POS tagging. We filter out adjectives and adverbs, storing them in a word set $\textsf{SLIM}_{\textsc{pos}}$. Finally, we perform a subset operation on $\textsf{SLIM}_{\textsc{pos}}$ to extract varying proportions of words. Specifically, after obtaining $\textsf{SLIM}_{\textsc{pos}}$, we extract the top $k$ proportion of words based on their indices, where $k$ corresponds to the desired proportion of the total word count. For NER tagging, the tokenization process is similar to POS tagging. After obtaining tokens, we use the \textit{ne\_chunk} function to extract the filtered named entities, which are then stored in words set $\textsf{SLIM}_{\textsc{ner}}$. We do not perform additional operations and restrictions for NER words since the named entities in an article are generally not too many, such as a person, location, and the like. 

\subsubsection{$\textsf{SLIM}_{\textsc{metadata}}$}  
The input to $\textsf{SLIM}_{\textsc{metadata}}$ consists solely of metadata to explore whether metadata can replace lengthy texts as key information for fake news detection. The metadata contained in different datasets varies. In light of the aforementioned papers, we will focus on textual data such as \texttt{title} (which we denote as $\textsf{SLIM}_{\textsc{title}}$) and \texttt{author} ($\textsf{SLIM}_{\textsc{author}}$) rather than discrete data. \textbf{XLNet}$_{\text{base}}$ is used as the encoder to generate embeddings for metadata and other types of information.

\subsubsection{$\textsf{SLIM}_{\textsc{multimodal}}$}  
The input to $\textsf{SLIM}_{\textsc{multimodal}}$ involves integrations of different types of aforementioned inputs such as various percentages of keywords sets and NER words ($\textsf{SLIM}_{\textsc{keyword}} \oplus \textsf{SLIM}_{\textsc{ner}}$), as well as combinations of keywords sets and different types of metadata ($\textsf{SLIM}_{\textsc{keyword}} \oplus \textsf{SLIM}_{\textsc{metadata}}$). Formally,
\begin{equation}
\textsf{SLIM}_{\textsc{multimodal}}^{\uppercase\expandafter{\romannumeral1}}  = \textsf{SLIM}_{\textsc{keyword}} \oplus \textsf{SLIM}_{\textsc{ner}}
\end{equation}
\begin{equation}
\textsf{SLIM}_{\textsc{multimodal}}^{\uppercase\expandafter{\romannumeral2}}  = \textsf{SLIM}_{\textsc{keyword}} \oplus \textsf{SLIM}_{\textsc{author}}
\end{equation}
\begin{equation}
\textsf{SLIM}_{\textsc{multimodal}}^{\uppercase\expandafter{\romannumeral3}}  = \textsf{SLIM}_{\textsc{keyword}} \oplus \textsf{SLIM}_{\textsc{title}}
\end{equation}
where $\oplus$ is the concatenation operator. 
\newline\newline\textbf{Framework}: Given an input sequence $x$, we define its length as $T$ (the number of words). During the pre-training phase, although we employ \textbf{XLNet}$_{\text{base}}$ as our pre-training model, the pre-training objective function is indeed crucial. This is because it facilitates a deeper understanding of the semantic and structural relationships inherent within the text. Throughout the pre-training process, this objective function enables the model to discern between distinct categories of keyword combinations (e.g., real news versus fake news), which gives the downstream classification tasks more robust features. The pre-training objective function, as defined in Equation \ref{equ_object}, employs XLNet’s permutation language modeling to capture contextual information from the input.
\begin{equation} \label{equ_object}
\mathcal{F}(\theta) = \max_{\underset{\theta}{}}\mathbb{E}_{z \sim \mathcal{Z}_T} \left[ \sum_{t=1}^T \log p(x_{zt} \mid \mathbf{x}  _{z_{<t}}; \theta) \right]
,\end{equation} where in our case, $\mathbf{x}$ is the $\textsf{SLIM}_{\textsc{keyword}}$ (and other defined inputs), $\mathcal{Z}_T$ represents the set of permutations of keywords set of length $T$. We use $zt$ to represent the $t_{th}$ element in $\mathcal{Z}_T$, and $z_{<t}$ to represent the $1_{st}$ to $t-1$ elements of $z \in \mathcal{Z}_T$.

The likelihood function in equation \ref{equ_object} is defined as 
\begin{equation} \label{equ_7}
p_{\theta}(X_{z_t} = x|\mathbf{x}  _{z_{<t}}) = \frac{exp(e(x)^Tg_{\theta}(\mathbf{x}  _{z_{<t}},z_t))}{\sum_{x'}exp(e(x')^Tg_{\theta}(\mathbf{x}  _{z_{<t}},z_t))}
,\end{equation} where $g_{\theta}$ is the two-stream self-attention model. 
\newline\newline\textbf{Fake News Detection}:
Finally, we will conduct the downstream task, which is fake news detection. Building upon the aforementioned inputs, we will directly load the pre-trained weights of \textbf{XLNet}$_{\text{base}}$ model and fine-tune it using our defined \textsf{SLIM} variants. The loss function in the fine-tuning stage of the \textsf{SLIM} framework is the cross entropy loss.
\begin{equation} 
\mathcal{L_{\textsf{SLIM}}}(\theta) = \mathcal{L_{\textbf{CE}}}(\theta) = -\frac{1}{N} \sum_{i=1}^N \sum_{c \in \mathcal{C}} y_{i,c} \log p(y_{i,c} = 1 \mid x_i, \theta)
,\end{equation} where $N$ is the sample size, and $y$ is the label for the input words set. The parameter $\theta$ is updated by:
\begin{equation} 
\theta_{t+1} = \theta_t - \eta \cdot \frac{m_t}{\sqrt{v_t} + \epsilon}
,\end{equation} where $\eta$ is the learning rate and set to 5$\cdot10^{-5}$, $m_t$ is the momentum estimate, $v_t$ represents the squared gradient estimate, and $\epsilon$ is the stability constant and set to 1$\cdot10^{-8}$.
\newline For the prediction, the optimization target is to minimize the cross-entropy loss between the predicted logits from the fine-tuned \textbf{XLNet}$_{\text{base}}$ model and the ground-truth labels of the fake news detection task under the Adam optimizer. 

\subsection{Quantifying Limited Information}
In order to better quantify and compare the information density of limited information with that of the full text, we employed two methods. The first method targets information density: we have proposed a method based on Shannon entropy, which we refer to as \textit{normalized Shannon entropy} for fake news detection. The second method explores the relationship of average token counts, which not only provides a more intuitive representation of the difference in information volume between inputs but also illustrates that fewer tokens correspond to reduced costs for future, more extensive commercial language models.
\subsubsection{Normalized Shannon Entropy}
In information theory, Shannon entropy \cite{shannon1948} measures the average uncertainty of information and is defined  as:
\begin{equation}
H(X) = - \sum\limits_{x\in \chi}p(x)\log_2 p(x)    
,\end{equation}
where $p(x)$ is the probability of $x$ in the distribution $\chi$.
In the context of a news article $A = \{w_1, w_2, ... w_p\}$, $p(x)$ is modeled as the relative frequency of each word $w_i$ in the article. Words that appear more often in the article have a higher probability of being randomly chosen from the article. Thus, we define the significance level of a word $w$ as: 
\begin{equation}
sig(w) = \frac{{f_w}^{\mathbf{\mathcal{T}}}}{|\mathbf{\mathcal{T}}|}
,\end{equation}
where ${f_w}^{\mathbf{\mathcal{T}}}$ is the word frequency of $w$ within the original full text $\mathbf{\mathcal{T}}$, and $|\mathbf{\mathcal{T}}|$ represents the total number of words in the full text $\mathbf{\mathcal{T}}$. Thus, we can represent the information density by calculating the information score $S_\textit{normalized}$ under normalized Shannon entropy of an arbitrary article $A$ as:
\begin{equation}
S_\textit{normalized} = \sum\limits_{w\in A}\frac{H(w)}{sig(w)}
\end{equation}
\newline\newline\textbf{Mathematical Interpretation} By dividing Shannon entropy by significance level, we can obtain the average information uncertainty per unit of the significance level. This ratio helps to numericalize the information density of each unit of importance. Additionally, when the range of significance levels is broad (e.g., some words are very frequent while others are rare), dividing Shannon entropy by significance level helps to mitigate the scale effect, making the measure of information density more consistent.

\subsubsection{Average Token Counts}
Tokens serve as the building blocks of the original text, enabling the model to process and generate natural language in a structured way \cite{vaswani2017attention}. A fixed tokenizer aims to maintain a consistent informational value for each token, so a reduction in token count generally conveys less information and diminishes the expression of information. Hence, we calculated the average token count for different types of inputs and compared them with the token count of the full text to verify that our inputs are sparser.

\subsubsection{Information Density Comparisons}
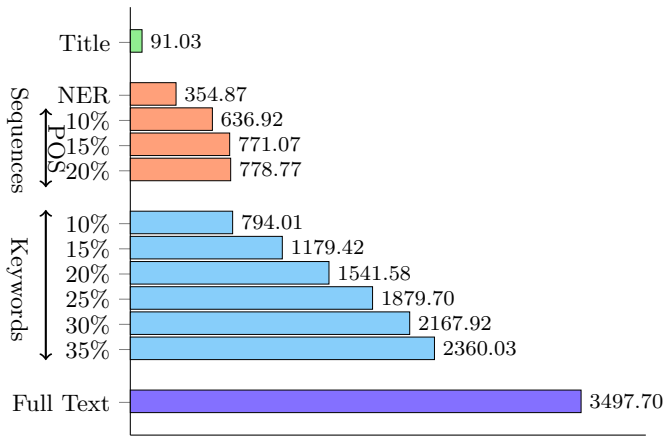
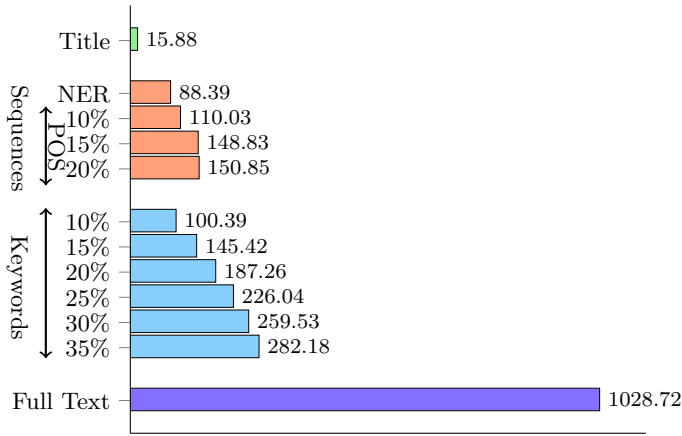
\begin{figure}[!t]
\begin{subfigure}{\columnwidth}
    \centering
  \begin{tikzpicture}
  \centering
    \begin{axis} [xbar,
                  bar width=0.3cm,
                  xtick=\empty,
                  ytick={0.3, 2.4, 3.4, 4.4, 5.4, 6.4, 7.4, 9.5, 10.5, 11.5, 12.5, 14.6},
                  xmin=0,
                  xmax=4000,
                  ymin=-1,
                  ymax=16,
                  nodes near coords,
                  point meta=explicit symbolic, 
                  every node near coord/.append style={font=\fontsize{7.7}{14}\selectfont, anchor=west}, nodes near coords align={center},
                  ylabel={},
                  xlabel={},
                  yticklabels={Full Text, , , , , , , , , , , Title},
                  axis y line*=left, 
                  axis x line*=bottom,
                  extra y ticks={2.4, 3.4, 4.4, 5.4, 6.4, 7.4, 9.5, 10.5, 11.5, 12.5}, 
                  extra y tick labels={35\%, 30\%, 25\%, 20\%, 15\%, 10\%, 20\%, 15\%, 10\%, NER}, 
                  extra y tick style={ticklabel style={anchor=east}, tick style={draw=none}},
                  clip=false,
                  ]

        \addplot [draw=black, line width=0.1mm, fill=LightSlateBlue] coordinates {(3497.70, 2) [3497.70]}; 
        
        \addplot [draw=black, line width=0.1mm, fill=LightSkyBlue] coordinates {
            (2360.03, 3) [2360.03] 
    	(2167.92, 4) [2167.92]
    	(1879.70, 5) [1879.70]
            (1541.58, 6) [1541.58]
            (1179.42, 7) [1179.42]
            (794.01, 8) [794.01]};
    
        \addplot [draw=black, line width=0.1mm, fill=LightSalmon] coordinates {
            (778.77, 9) [778.77]
            (771.07, 10) [771.07]
            (636.92, 11) [636.92]
            (354.87, 12) [354.87]};
    
        \addplot [draw=black, line width=0.1mm, fill=LightGreen] coordinates {
            (91.03, 13) [91.03]};

    \end{axis}
    \node[anchor=west, rotate=270] at ([yshift=4.75cm, xshift=-1.5cm] current axis.south west) {Sequences};
    \node[anchor=west, rotate=270] at ([yshift=4.25cm, xshift=-0.98cm] current axis.south west) {POS};
    \draw[<->, thick] ([yshift=0.45cm, xshift=-4.55cm] current axis.south|-current axis.west) -- ([yshift=1.5cm, xshift=-4.55cm] current axis.north|-current axis.west);

    \node[anchor=west, rotate=270] at ([yshift=2.75cm, xshift=-1.5cm] current axis.south west) {Keywords};
    \draw[<->, thick] ([yshift=-1.85cm, xshift=-4.55cm] current axis.south|-current axis.west) -- ([yshift=0.15cm, xshift=-4.55cm] current axis.north|-current axis.west);
    
    \end{tikzpicture}
  \caption{Representation of information density by average normalized Shannon entropy ( ${\bar{S}}_{normalized}$) on the \textsc{ReCOVery} dataset. The Title shows the lowest normalized Shannon score (91.03). NER words also have lower Shannon scores (\textasciitilde 10\% of full text density). Keywords and POS words at 10\% threshold show significantly lower Shannon scores than full text} 
  \label{shannon_entropy} 
\end{subfigure}
\vspace{1cm}  
\begin{subfigure}{\columnwidth}
    \centering
  \begin{tikzpicture}
    \begin{axis} [xbar,
                  bar width=0.3cm,
                  xtick=\empty,
                  ytick={0.3, 2.4, 3.4, 4.4, 5.4, 6.4, 7.4, 9.5, 10.5, 11.5, 12.5, 14.6},
                  xmin=0,
                  xmax=1130,
                  ymin=-1,
                  ymax=16,
                  nodes near coords,
                  point meta=explicit symbolic, 
                  every node near coord/.append style={font=\fontsize{7.7}{14}\selectfont, anchor=west}, nodes near coords align={center},
                  ylabel={},
                  xlabel={},
                  yticklabels={Full Text, , , , , , , , , , , Title},
                  axis y line*=left, 
                  axis x line*=bottom,
                  extra y ticks={2.4, 3.4, 4.4, 5.4, 6.4, 7.4, 9.5, 10.5, 11.5, 12.5}, 
                  extra y tick labels={35\%, 30\%, 25\%, 20\%, 15\%, 10\%, 20\%, 15\%, 10\%, NER}, 
                  extra y tick style={ticklabel style={anchor=east}, tick style={draw=none}},
                  ]

        \addplot [draw=black, line width=0.1mm, fill=LightSlateBlue] coordinates {(1028.72, 2) [1028.72]}; 
        
        \addplot [draw=black, line width=0.1mm, fill=LightSkyBlue] coordinates {
            (282.18, 3) [282.18]
    	(259.53, 4) [259.53]
    	(226.04, 5) [226.04]
            (187.26, 6) [187.26]
            (145.42, 7) [145.42]
            (100.39, 8) [100.39]};
    
        \addplot [draw=black, line width=0.1mm, fill=LightSalmon] coordinates {
            (150.85, 9) [150.85] 
            (148.83, 10) [148.83]
            (110.03, 11) [110.03]
            (88.39, 12) [88.39]};
    
        \addplot [draw=black, line width=0.1mm, fill=LightGreen] coordinates {
            (15.88, 13) [15.88]}; 

    \end{axis}
    
    \node[anchor=west, rotate=270] at ([yshift=4.75cm, xshift=-1.5cm] current axis.south west) {Sequences};
    \node[anchor=west, rotate=270] at ([yshift=4.25cm, xshift=-0.98cm] current axis.south west) {POS};
    \draw[<->, thick] ([yshift=0.45cm, xshift=-4.55cm] current axis.south|-current axis.west) -- ([yshift=1.5cm, xshift=-4.55cm] current axis.north|-current axis.west);

    \node[anchor=west, rotate=270] at ([yshift=2.75cm, xshift=-1.5cm] current axis.south west) {Keywords};
    \draw[<->, thick] ([yshift=-1.85cm, xshift=-4.55cm] current axis.south|-current axis.west) -- ([yshift=0.15cm, xshift=-4.55cm] current axis.north|-current axis.west);
    
    \end{tikzpicture}
  \caption{Representation of information density by the average count of tokens on the \textsc{ReCOVery} dataset. The title and NER words have the lowest average token count among all types. All inputs, including sequences and keywords at different percentages, have much lower token counts than full text, with the highest reaching only \textasciitilde27\% of full text length.} 
  \label{tokencount}   \end{subfigure}
\vspace{-45pt}\caption{Representation of information density by average normalized Shannon entropy (a) and the average count of tokens (b) on the \textsc{ReCOVery} dataset}
\end{figure}

To compare whether the different types of input we designed in section \ref{formulation} indeed contain limited and less information, we calculated the information density of each type using the average token count and proposed normalized Shannon entropy score. The results on the ReCOVery dataset, presented in Figure \ref{shannon_entropy} and \ref{tokencount} separately, reveal the following: the title exhibits the lowest normalized Shannon score (91.03) and count of tokens (15.88) due to its inherent conciseness as part of the metadata. NER words, as an effective representation for identifying and classifying key entities, also show a low score of 354.87, which is 10\% of the full text, and token counts of 88.39, 8.59\% of the full text. Additionally, both POS words and keywords, with the default 10\% proportion, demonstrate significantly lower Shannon scores and token counts compared to the full text. It is noteworthy that both information density evaluation metrics for POS words do not exhibit a linear increase as the percentage rises. The figures of normalized Shannon entropy score and average token count for the remaining two datasets are presented in Appendix \ref{appendix_info_density}.

\subsection{Fake News Detection}\label{Phase3}
Ultimately, we will conduct the downstream task, which is fake news detection. Building upon the aforementioned framework formulation, we will employ language models for fake news detection, as language model-based approaches currently yield the best performance for detecting fake news. 
\subsubsection{Base configurations} 
We will use \textbf{XLNet}$_{\text{base}}$ as the encoder to generate the corresponding embeddings of the input information \cite{yang2019xlnet}. We use Adam in the optimization process. For the prediction phase, we apply the \texttt{argmax} function to the logits from XLNet to obtain the final prediction label. Mathematically, 

\begin{equation}
\hat{y}= \operatorname{argmax}_{i}(\mathbf{z}_i),
\end{equation}
where $\hat{y}$ is the predicted label and $\mathbf{z}_i$ is the logits computed by the final layer of the \textsf{SLIM} framework.

\begin{table*}[t] 
    \centering 
    \caption{Performance comparison of datasets on the SLIM, CapsNet, MisROB\AE RTA, and selected DocEmb models. The percentage of keywords used in comparisons for both types is 25\%. The best performance is highlighted in bold, and the second best is underlined.} 
    \renewcommand{\arraystretch}{1.1} 
    \small 
    \begin{adjustbox}{width=0.7\textwidth}
    \begin{tabular}{lcc} 
        \hline 
        \multirow{2}{*}{Method} & \multicolumn{2}{c}{Dataset} \\
        \cline{2-3} & ReCOVery & Fake\_And\_Real\_News \\ 
        \hline
        DocEmb\_{\tiny TFIDF} BiLSTM
            & 89.56$\pm$0.0025
            & 92.26$\pm$0.0032 \\
        DocEmb\_{\tiny TFIDF}  BiGRU
            & 90.54$\pm$0.0017 
            & 92.60$\pm$0.0028 \\
        DocEmb\_{\tiny BERT} BiLSTM
            & 90.27$\pm$0.0033  
            & 93.05$\pm$0.0026\\
        DocEmb\_{\tiny BERT} BiGRU
            & 90.13$\pm$0.0014 
            & 93.07$\pm$0.0051 \\
        \hdashline
        MisROB\AE RTA
            & 91.35$\pm$0.0066 
            & \underline{97.34$\pm$0.0076} \\
        BiLSTM\_CapsNet
            & \underline{95.49$\pm$0.0134}
            & 95.56$\pm$0.0091 \\
        \hdashline    
        \textsf{SLIM}           
            & \scalebox{0.93}{\textbf{95.55$\pm$0.0046}}
            & \scalebox{0.93}{\textbf{97.60$\pm$0.0031}}    \\
        $\textsf{SLIM}_{\textsc{keyword}}$
            &92.86$\pm$0.0070
            &92.76$\pm$0.0016  \\
        $\textsf{SLIM}_{\textsc{multimodal}}^{\uppercase\expandafter{\romannumeral3}}$
            & 93.72$\pm$0.0074
            & 93.72$\pm$0.0049   \\\hline
    \end{tabular}
    \end{adjustbox}
    \label{table_result_compare} 
\end{table*}

      
         
      
       
       
       
        
        
       

\section{Experimental Results}\label{expe}
In this section, we will introduce the experimental setup, including preprocessing and datasets. Subsequently, we conducted extensive experiments to address the following five research questions, RQ1 through RQ5. The research questions are as follows: \textbf{RQ1}: How does \textsf{SLIM} compare to other baselines? \textbf{RQ2}: How effective are keywords for fake news detection? \textbf{RQ3}: How effective are sequences for fake news detection? \textbf{RQ4}: How effective is metadata for fake news detection? \textbf{RQ5}: Can multiple modalities of limited information enhance fake news detection?
\subsection{Experimental Setup}
For each experiment, we conducted five trials to obtain the average accuracy. During data preprocessing, paragraph separators `$\backslash n$' were removed, and all text was converted to lowercase to ensure consistency.
\subsubsection{Dataset}
Our experiments are conducted on two public benchmark datasets of fake news detection: ReCOVery \cite{zhou2020recovery}, and Fake\_And\hyp{}Real\hyp{}News \cite{fake_real_news_dataset}. The division of training, validation, and testing sets in the ReCOVery are in the same way as the articles from which they are derived. The training, validation, and testing sets are divided in a ratio of $50\%: 25\%: 25\%$ in the Fake\_And\_Real\_News dataset. The basic statistics of the datasets and detailed source descriptions of these datasets are in Table \ref{table: datasetstat} and Appendix \ref{dataset_des}.
\begin{table}[t]
\caption{Dataset statistics}\vspace{-3mm}
\centering
\renewcommand{\arraystretch}{1.2} 
\Large  
\resizebox{0.98\columnwidth}{!}{ 
\begin{tabular}{l|l|rcc}
    \hline \hline
    Dataset & Labels & Train & Validation & Test\\
    \hline
   \multirow{2}*{\textsc{ReCOVery}} & Truth  & 966  & 278 & 120 \\
       & Fake & 487  & 114 & 64\\
    \hline
     \multirow{2}*{\textsc{Fake\_Real\_News}} & Truth  & 1143 & 557 & 597\\
       & Fake & 1154 & 592 & 551 \\
     \hline \hline
\end{tabular}}
\label{table: datasetstat}
\end{table}

\subsubsection{Metadata Selection}
The metadata we selected to use in our work contains textual data only. To be specific, \texttt{title, author} are selected in the ReCOVery dataset. Meanwhile, \texttt{title} is selected in the Fake\_And\_Real\_News dataset. Additionally, only \texttt{author} is selected in the ReCOVery dataset. 

\subsubsection{Evaluation Metrics}
We report accuracy, macro-$F_1$, and AUC. We also conduct statistical significance comparisons between different experimental groups. We use $^{**}$ to represent $p$-values below 0.01 and use $^*$ to represent $p$-values between 0.01 and 0.05 for two groups. The absence of asterisks indicates that there is no statistically significant difference between the two experimental groups.

\subsection*{RQ1: How does \textsf{SLIM} compare to other baselines?}
In this section, we present a comprehensive comparative analysis between our proposed \textsf{SLIM} framework against various state-of-the-art models, including different deep learning models and large language models. The baseline models we employed are described as follows.
\newline 
$\blacktriangleright$\textbf{DocEmb}: DocEmb was proposed by Truic{\u{a}} and Apostal \cite{truicua2023s}. Instead of relying on handcrafted features or complex deep learning architectures, the approach utilizes pre-trained document embeddings to capture the semantic meaning of news articles. These embeddings are then fed into models of neural network architecture. Based on the combinations with good performance presented in their paper, we utilize 4 different combinations in our work: 2 vectorization methods (TF-IDF, BERT) combined with 2 downstream neural network models (BiLSTM and BiGRU). 
\newline
$\blacktriangleright$ \textbf{BiLSTM\_Capsnet}: BiLSTM\_Capsnet was proposed by Sridhar and Sanagavarapu \cite{sridharcapsnet}. The framework uses a multi-task learning architecture. The architecture's subtasks include modeling the article contents, and the shared common task is determining whether or not the article is fake. The BiLSTM network is used to model the subtasks, and CapsNet serves as the common meta classifier.
\newline
$\blacktriangleright$\textbf{MisROB\AE RTA}:  MisROB\AE RTA was proposed by Truic{\u{a}} and Apostal \cite{truicua2022misrobaerta}. The model incorporates various techniques, such as data augmentation and adversarial training, to improve its robustness in detecting misleading content.

We first conducted experiments and obtained our baseline results of the datasets under the \textsf{SLIM} framework. The baseline entails using only the full-text body as input to build the XLNet model for prediction accuracy. The results of the \textsf{SLIM} baseline are presented in Table \ref{table_baseline}. 
We observed that the full text exerts heterogeneous impacts, however, the prediction accuracy for all datasets exceeded 93\%.

The comparison of the performance of different baselines is shown in Table \ref{table_result_compare}. The results illustrate that, compared to other baseline models, the \textsf{SLIM} achieved the highest accuracy in both the ReCOVery and Fake\_And\_Real\_News dataset. Meanwhile, by using only keywords with half the information density of the full text, we are able to achieve impressive accuracy. Not only does this performance closely approach some state-of-the-art fake news detection models (e.g., MisROB\AE RTA), but it also surpasses many of the latest deep learning and language model-based approaches (e.g., DocEmb). Moreover, when we combine keywords with the title (which always has the lowest information density), the accuracy is further improved.

\begin{table*}[t]
    \caption{Performance comparison of datasets of the \textsf{SLIM} (full-text) baseline frameworks: The performance of all datasets in fake news detection using the \textsf{SLIM} framework exceeded 93\%.}
    \vspace{-3mm}
    \renewcommand{\arraystretch}{1.2} 
    \centering
    \small
    \begin{adjustbox}{width=0.8\textwidth}
    \begin{tabular}{l|ccc|ccc}
        \hline
        \hline
        \multirow{2}{*}{Experiments} & \multicolumn{3}{c|}{ReCOVery}
        & \multicolumn{3}{c}{Fake\_And\_Real\_News}\\
        \cline{2-7}
        & Accuracy& Macro-$F_1$ & AUC & 
        Accuracy & Macro-$F_1$ & AUC\\
        \hline
        \textsf{SLIM}               
        & 95.55$\pm$0.0046   & 94.71   & 95.53
        & 97.60$\pm$0.0031   & 97.60   & 97.62  \\
        \hline
        \hline
    \end{tabular}
    \end{adjustbox}
    \label{table_baseline}
\end{table*}

\subsection*{RQ2: How effective are keywords for fake news detection?}

Subsequently, we primarily investigated the impact of limited yet effective information (except metadata) mentioned in the first two phases of section \ref{methodology} on the $\textsf{SLIM}_{\textsc{keyword}}$ framework. Initially, we explored the effect of keywords on fake news detection. We extracted keyword sets from different datasets using the methodology outlined in section \ref{kwextract}. Additionally, for each dataset, we attempted to extract the maximum percentage of keywords feasible (rounded down using the floor function). We set the default, i.e., the minimum percentage of keywords, to be 10\% of the original full text. Then, for each dataset, we gradually increased the percentage of keywords extracted by 5\% for experimentation. The results are depicted in Figure \ref{fig_keywords_only}.
From Figure \ref{fig_keywords_only}, we set the y-axis as the prediction accuracy divided by the baseline accuracy (referred to as the \textit{accuracy ratio}), as this provides a more intuitive way to visualize the impact of keywords on detection from both the graphical and numerical perspectives. The following figures utilize this y-axis configuration.

In summary, for all datasets, there is an overall trend of increasing accuracy ratio as the percentage of keywords increases. Across all datasets except for the Fake\_And\_Real\_News dataset, once the extracted keywords reach 30\% of the text, we observe that the accuracy ratio reaches approximately 99\%. This indicates that comparable and good performance can be achieved by extracting only 30\% of the full text, significantly reducing computational inefficiency and enhancing scalability for large datasets. This finding implies that keyword extraction can effectively filter out irrelevant words and information in fake news detection.
\begin{figure}[t]
\centering
\begin{tikzpicture}
  \begin{axis}[ 
  width=0.95\linewidth, 
  height=0.7\linewidth, 
  line width=0.5,
  grid=major, 
  ymajorgrids=true, 
  xmajorgrids=false, 
  tick label style={font=\normalsize},
  label style={font=\normalsize},
  xlabel={Percentage of Keywords},
  ylabel={Prediction Accuracy/Baseline},
  ylabel style={yshift=-7pt}, 
  y tick label style={
    /pgf/number format/.cd,
    fixed,
    fixed zerofill,
    precision=2
  },
  legend style={at={(0,1)}, anchor=north west, draw=black, fill=white, nodes={scale=0.7, transform shape}},
  axis line style={draw=black}, 
  ymin=0.92,
  ymax=1.01,
  xtick pos=left, 
  ytick pos=left, 
  ]
    \addplot[blue, mark=square*, mark size=1.3] coordinates
      {(10,0.9422) (15,0.9541) (20,0.9614) (25,0.9714) (30,0.9879) (35,0.9888)};
      \addlegendentry{ReCOVery}
    
    \addplot[orange, mark=*, mark size=1.5] coordinates
      {(10,0.9268) (15,0.9417) (20,0.9459) (25,0.9504) (30,0.9599) (35,0.9641) (40,0.9708)};
    \addlegendentry{Fake\_And\_Real\_News} 
  \end{axis}
\end{tikzpicture}\vspace{-3mm}
\caption{Performance comparison of datasets of the $\textsf{SLIM}_{\textsc{keyword}}$ frameworks. All datasets achieve an accuracy ratio of over 96\% when we extract 30\% of the keywords, among which the ReCOVery datasets showed an approximately 99\% accuracy ratio.}
\label{fig_keywords_only}\vspace{-2mm}
\end{figure}
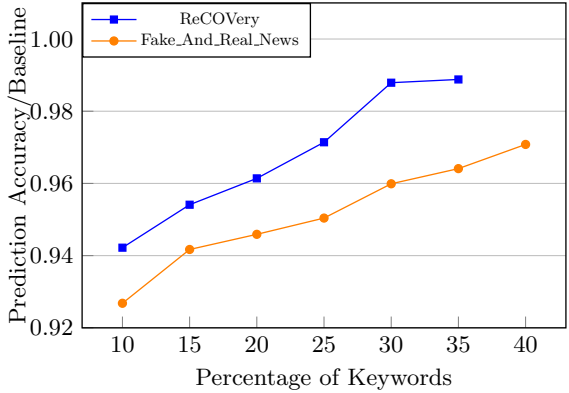

 \begin{figure}[t]
\centering
\begin{tikzpicture}
  \begin{axis}[ 
  width=0.95\linewidth, 
  height=0.7\linewidth, 
  line width=0.5,
  grid=major, 
  ymajorgrids=true, 
  xmajorgrids=false, 
  tick label style={font=\normalsize},
  label style={font=\normalsize},
  xlabel={Percentage of POS tagging words},
  ylabel={Prediction Accuracy/Baseline},
  ylabel style={yshift=-7pt}, 
  y tick label style={
    /pgf/number format/.cd,
    fixed,
    fixed zerofill,
    precision=2
  },
  xtick={10, 15, 20}, 
  legend style={at={(0,1)}, anchor=north west, draw=black, fill=white, nodes={scale=0.7, transform shape}},
  axis line style={draw=black}, 
  ymin=0.93,
  ymax=0.955,
  xtick pos=left, 
  ytick pos=left, 
  ]
    \addplot[blue, mark=square*, mark size=1.3] coordinates
      {(10,0.9395) (15,0.9422) (20,0.9514)};
      \addlegendentry{ReCOVery}
    
    \addplot[orange, mark=*, mark size=1.5] coordinates
      {(10,0.9300) (15,0.9388)};
      \addlegendentry{Fake\_And\_Real\_News}
  \end{axis}
\end{tikzpicture}\vspace{-4mm}
\caption{Performance comparison of datasets of the $\textsf{SLIM}_{\textsc{sequence}}$ frameworks in POS tagging words.The percentage of POS tagging words (primarily adjectives and adverbs) that can be extracted from the full text is approximately 10\% to 20\%. However, using a small number of POS tagging words can achieve an accuracy ratio of 94\%.}
\label{fig_postagging}\vspace{-9pt}
\end{figure}
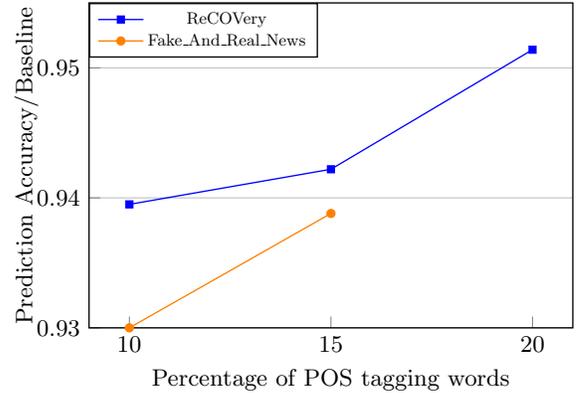
\subsection*{RQ3: How effective are sequences for fake news detection?}
Within the $\textsf{SLIM}_{\textsc{sequence}}$ framework, we also explored the impact of POS tagging words and NER tagging words on fake news detection. For POS tagging words, adjectives and adverbs are particularly powerful for enhancing fake news detection, given their frequent usage in texts to express authors' opinions and emotions. Therefore, we adopted a similar approach to extracting the percentage of POS tagging words as with keywords. As for NER words, since the occurrence of named entities in texts is not typically abundant, we did not impose any percentage limitations during extraction. Our experiments demonstrated that, across the two datasets, NER tagging words accounted for approximately 10\%, which is consistent with our default minimum percentage. The results of $\textsf{SLIM}_{\textsc{sequence}}$ framework regarding POS tagging and NER tagging are presented in Figure \ref{fig_postagging} and Table \ref{table_ner_tag}, respectively. We can observe from Figure \ref{fig_postagging} that the maximum percentage of POS tagging words that can be extracted from the Fake\_And\_Real\_News datasets is 15\%. Meanwhile, the ReCOVery dataset allows for the extraction of up to 20\% of the POS tagging words from the original text. As a result, POS tagging shows an overall increasing trend across all datasets, where the accuracy ratio increases as the percentage of POS tagging words increases. However, compared to the performance of keywords, the accuracy ratio of POS tagging words remains around 94\%.

Secondly, regarding the NER tagging words performance, in the ReCOVery dataset, NER tagging words achieve an 86.82\% accuracy (which is significantly lower than the baseline accuracy) and an accuracy ratio of 93\%. The prediction accuracy for the Fake\_And\_Real\_News dataset is 90.08\%, with \textit{p}-value between 0.01 and 0.05, indicating a significant decrease compared to the baseline.

\subsection*{RQ4: How effective is metadata for fake news detection?}
In practical scenarios, we often observe a partial overlap between the information contained in metadata (such as \texttt{title}) and the content of the text body \cite{piotrkowicz2017automatic}. As a result, the overlapped information is redundantly utilized during tokenization, leading to reduced efficiency and increased consumption of embedding resources. Therefore, we aim to mitigate the drawbacks mentioned before. As metadata usually contains the minimum of information needed to distinguish an article, we aim to explore whether fake news detection can be achieved only through metadata, replacing the need for the full-text body. We exclusively use metadata as the input, feeding it directly into the $\textsf{SLIM}_{\textsc{metadata}}$  framework to obtain the results. To be more precise, for the ReCOVery dataset, its metadata includes both \texttt{author} and \texttt{title}. Therefore, we input these two pieces of metadata separately to obtain the results. However, for the Fake\_And\_Real\_News dataset, its metadata only includes the \texttt{title}. Hence, the input is the \texttt{title}. The results are in Table \ref{table_metadata}.

\begin{table}[t]
    \caption{Performance comparison of datasets of the $\textsf{SLIM}_{\textsc{sequence}}$ frameworks in NER tagging words. The performance of NER words exhibits heterogeneous effects across different datasets} \vspace{-3mm}
    \centering
    \renewcommand{\arraystretch}{1.1} 
    \LARGE
    \begin{adjustbox}{width=1.0\columnwidth} 
    \begin{tabular}{l|ccc}
        \hline \hline
        \multirow{2}{*}{Dataset}  & \multicolumn{3}{c}{$\textsf{SLIM}_{\textsc{ner}}$} \\
        \cline{2-4}
        & Accuracy& Macro-$F_1$ & AUC \\
        \hline
        \textsc{ReCOVery}      &86.82$^{**}\pm$0.0078   & 83.78   & 83.29\\
        \hline
        \textsc{Fake\_And\_Real\_News}   &90.08$^{*}\pm$0.0092   & 90.08  & 90.14\\
        \hline
        \hline 
    \end{tabular}
    \end{adjustbox}
    \label{table_ner_tag}
    \vspace{-2pt}
\end{table}

We discover that, from Table \ref{table_metadata}, utilizing only textual metadata (\texttt{title} and \texttt{author} in this work) as input for the fake news detection results in a statistically significant decrease in prediction accuracy compared to the baseline (which uses the full-text body as the input) performance. Specifically, in the ReCOVery and Fake\_And\_Real\_News dataset, when using metadata alone as a single input for detection under the $\textsf{SLIM}_{\textsc{metadata}}$  framework, the accuracy generally decreases by approximately 10\% compared to the baseline. Without considering any text, we could not achieve the same level of accuracy by exclusively using metadata for fake news detection. However, if aiming for a relatively good level of accuracy, we can use metadata or selectively combine less information of full text for future fake news detection.

\begin{table}[t]
    \caption{Performance of the metadata-only framework ($\textsf{SLIM}_{\textsc{metadata}}$). Metadata cannot substitute text, yielding results significantly lower to the results obtained using text alone.} \vspace{-3mm}
    \centering
    \renewcommand{\arraystretch}{1.1} 
    \LARGE
    \begin{adjustbox}{width=1.0\columnwidth} 
    \begin{tabular}{l|ccc}
        \hline
        \hline
        \multirow{2}{*}{Dataset} & 
        \multicolumn{3}{c}{$\textsf{SLIM}_{\textsc{metadata}}$}\\
        \cline{2-4}
        & Accuracy& Macro-$F_1$ & AUC \\
        \hline
        \textsc{ReCOVery (\texttt{title})}      
        &82.25$^{**}\pm$0.0066   & 78.14   & 77.80  \\
        \hline
         \textsc{ReCOVery (\texttt{author})}      
         &76.99$^{**}\pm$0.0071 & 74.54  &77.62 \\
        \hline
        \textsc{Fake\_And\_Real\_News (\texttt{title})} 
        &85.21$^{**}\pm$0.0034  & 85.19  & 85.42  \\
        \hline
        \hline 
    \end{tabular}
    \end{adjustbox}   
    \label{table_metadata} 
    \vspace{-3pt}
\end{table}

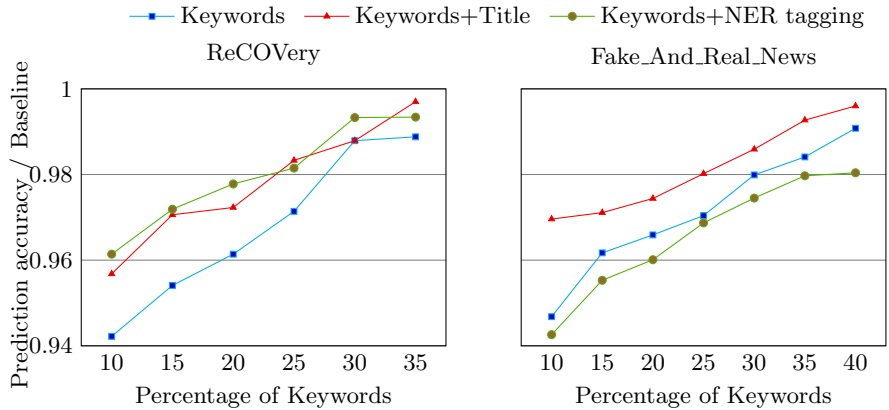
\begin{figure*}[t!]
\centering
\begin{tikzpicture}
  \begin{groupplot}[
      group style={
          group name=my plots,
          group size=2 by 1,
          xlabels at=edge bottom,
          ylabels at=edge left,
          horizontal sep=1cm,
          vertical sep=1cm,
          y descriptions at=edge left, 
      },
      ylabel={Prediction accuracy / Baseline},
      legend style={
          at={(-0.06,1.2)},        
          anchor=south,           
          legend columns=3,       
          /tikz/every even column/.append style={column sep=5pt}, 
          draw=none               
      },
      width=0.363\textwidth,
      height=5cm,                 
      grid=major,
      grid style={gray},
      xmajorgrids=false,
      yminorticks=false,
      xlabel near ticks,
      ylabel style={xshift=-3pt, yshift=-8pt},
      xtick style={draw=none},
      ytick style={draw=none}
  ]

  \nextgroupplot[title=ReCOVery, xlabel=Percentage of Keywords, ymin=0.94, ymax=1, xtick={10,15,...,35}]
  \addplot+[cyan, mark=square*, mark size=1.1] coordinates {(10,0.9422)(15,0.9541)(20,0.9614)(25,0.9714)(30,0.9879)(35,0.9888)};
  \addplot+[red, mark=triangle*, mark size=1.3] coordinates {(10,0.9568)(15,0.9706)(20,0.9723)(25,0.9833)(30,0.9879)(35,0.9970)};
  \addplot+[red!30!green, mark=*, mark size=1.5] coordinates {(10,0.9614)(15,0.9719)(20,0.9778)(25,0.9815)(30,0.9933)(35,0.9934)};
  
  \nextgroupplot[title=Fake\_And\_Real\_News, xlabel=Percentage of Keywords, ymin=0.92, ymax=0.98, xtick={10,15,...,40}, ytick={0.92,0.94,...,0.98}]
  \addplot+[cyan, mark=square*, mark size=1.1] coordinates {(10,0.9268)(15,0.9417)(20,0.9459)(25,0.9504)(30,0.9599)(35,0.9641)(40,0.9708)};
  \addplot+[red, mark=triangle*, mark size=1.3] coordinates{(10,0.9496)(15,0.9511)(20,0.9544)(25,0.9602)(30,0.9659)(35,0.9727)(40,0.9760)};
  \addplot+[red!30!green, mark=*, mark size=1.5] coordinates {(10,0.9226)(15,0.9353)(20,0.9401)(25,0.9487)(30,0.9545)(35,0.9597)(40,0.9604)};

  \legend{Keywords, Keywords+Title, Keywords+NER tagging}
  \end{groupplot}
\end{tikzpicture}
\caption{Performance comparison of datasets of the $\textsf{SLIM}_{\textsc{multimodal}}$ frameworks. Generally, the integration of different types of limited information improves fake news detection accuracy compared to using only keywords ($\textsf{SLIM}_{\textsc{keyword}}$). In the Fake\_And\_Real\_News dataset, the performance of keywords and NER words shows an approximately 0.5\% decline compared to using only keywords.} \vspace{-1mm}
  \label{keywords_type3_results} 
\end{figure*}

\subsection*{RQ5: Can multiple modalities of limited information enhance fake news detection?}
In this section of the experiment, we aim to investigate whether combining different pieces of limited key information can enhance the performance of the $\textsf{SLIM}_{\textsc{multimodal}}$  framework. Initially, for each dataset, we combined their respective percentages of keywords and NER tagging words. As mentioned in the methodology, we concatenated these two distinct word sets together to form a composite input for the encoder. The final results are depicted in Figure \ref{keywords_type3_results}. Additionally, we sought to integrate keyword information with metadata to assess whether metadata could serve as additional information to enhance the performance. The results are also presented in Figure \ref{keywords_type3_results}.

The results in Figure \ref{keywords_type3_results} lead us to the following conclusions. Firstly, in the ReCOVery dataset, we found that the integration of limited information $\textsf{SLIM}_{\textsc{multimodal}}$: keywords + \texttt{title}, keywords + NER tagging words) improves detection performance compared to using only keywords for fake news detection. Furthermore, we observed that NER words have a greater impact on fake news detection than metadata (\texttt{title}). Finally, in the Fake\_And\_Real\_News dataset, metadata can still be experimentally verified as useful for improving accuracy when combined with keywords. However, it is worthily noted that the heterogeneous effects of NER tagging exist, such that combining keywords with NER words results in a slight accuracy reduction of approximately 0.5\% compared to the $\textsf{SLIM}_{\textsc{keyword}}$.

\section{Conclusion and Future Work}
\sloppy In this work, we systematically investigated the viability of limited-information strategies for fake news detection using the \textsf{SLIM} framework. We investigated and conducted extensive experiments with different types of information strategies: keyword extraction, sequence tagging, and textual metadata. Our empirical analysis demonstrates that strategic keyword extraction preserves critical information even under severe sparsity constraints: retaining merely 30\% of full-text keywords achieves a near-perfect accuracy ratio (99\%) across multiple benchmarks. Linguistic tagging experiments further revealed that limited syntactic-semantic representations suffice for detection. Constrained POS and NER tagging sets independently achieved a 92\% accuracy ratio. While metadata exhibited diminished standalone performance, its complementary role in the multimodal framework proved statistically significant. Our systematic evaluation of multi-modality limited information demonstrates that multi-view fusion of keywords, named entities, or contextual titles achieves substantial performance increase: not only does this combination surpass single-modality keyword analysis, but it also consistently outperforms state-of-the-art neural network approaches across two benchmark datasets. Our findings substantiate that strategically selected information subsets can achieve accuracy parity with full-text analysis, establishing an efficiency-optimized framework for fake news detection and providing guidelines for sparse-data environments where full-text acquisition is impractical. Future work will focus on enhancing robustness through syntactic-semantic augmentation techniques, including controlled paraphrase generation and dependency shuffling.


\section{Acknowledgements}
This research was supported in part by the National Science Foundation under award  No. 2241070.

%
\bibliographystyle{abbrv}
\bibliography{sample-base}  
%
%
\appendix
\begin{figure}[t]
  \centering
  \begin{subfigure}[t]{\linewidth}
    \centering
    \includegraphics[width=0.9\linewidth]{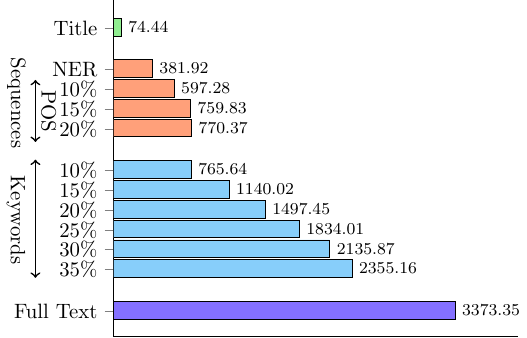}
    \caption{Representation of information density by average normalized Shannon entropy on the Fake\_And\_Real\_News dataset. The title yields the lowest score of 74.44. NER words exhibit a relatively lower Shannon score, capturing 11\% of the information density of the full text. Similarly, both keywords and POS words, when sampled at the default 10\%, demonstrate significantly lower scores compared to the full text} 
    \label{fig:fnrshannon}
  \end{subfigure}
\vspace{-2mm}
  \begin{subfigure}[t]{\linewidth}
    \centering
    \includegraphics[width=0.9\linewidth]{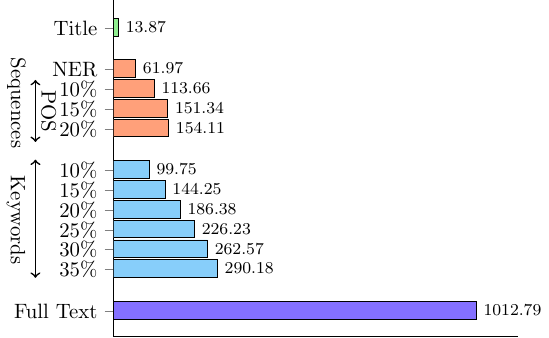}
    \caption{Representation of information density by the average count of tokens on the Fake\_And\_Real\_News dataset. The title and NER words maintain the lowest token counts among all input types. Across all inputs, the token counts remain significantly lower than those of the full text, with the highest reaching about 29\% of the full text.} 
    \label{fig:fnrtoken}
  \end{subfigure}
\vspace{-1mm}
  \caption{Representation of information density by average normalized Shannon entropy (a) and the average count of tokens (b) on the Fake\_And\_Real\_News dataset.}\vspace{-3.5mm}
\end{figure}

\section{Dataset}\label{dataset_des}
\subsection{ReCOVery Dataset}
The ReCOVery dataset is a repository that has been built to make it easier to conduct research on countering COVID-19-related information. After conducting a thorough search and investigation of around 2,000 news publishers, 60 were found to have extremely high or low levels of credibility by the authors of the dataset. The repository includes 2,029 news pieces about the coronavirus that were published between January and May 2020, as well as 140,820 tweets that show how these stories were shared on the Twitter social network. ReCOVery has a wide collection of news articles, social media posts, images, videos, and audio recordings pertaining to COVID-19. The dataset covers various themes and topics related to the pandemic, including public health guidance, government policies, scientific research, and societal impacts. Additionally, ReCOVery includes metadata such as publication dates, image, country, sources, and contextual information \cite{zhou2020recovery}.

Descriptions of the variables: \textit{\textbf{label}}: news label (1 = real, 0 = fake);  \textit{\textbf{text}}: content of the news; \textit{\textbf{title}} and \textit{\textbf{author}}.

\subsection{Fake\_And\_Real\_News Dataset}
The Fake\_And\_Real\_News Dataset comprises two distinct components sourced through different methods. The first part consists of 13,000 articles labeled as "fake news," obtained from a dataset released by Kaggle during the 2016 election cycle. For the second part,  To gather these, the author turned to All Sides, a platform hosting news and opinion pieces spanning the political spectrum. With articles categorized by topic and political leaning, All Sides facilitated web scraping from diverse media outlets, including prominent names like the New York Times, WSJ, Bloomberg, NPR, and the Guardian. Finally, a total of 5,279 real news articles published in 2015 or 2016 were successfully scraped. The dataset was meticulously constructed to ensure balance, with an equal number of fake and real articles, resulting in a null accuracy of 50\%. The finalized dataset encompasses 10,558 articles, complete with headlines, full-body text, and corresponding labels denoting their authenticity (real or fake) \cite{fake_real_news_dataset}. The dataset is publicly available in the provided GitHub repository.

Descriptions of the variables: \textit{\textbf{label}}: news label;  \textit{\textbf{text}}: content of the news; and \textit{\textbf{title}}.

\section{Information Density Comparisons}\label{appendix_info_density}
This section presents the remaining graphs for the representation of information density and average token count respectively for the Fake\_And\_Real\_News Dataset. Notably, the difference between full-text and keywords is significantly lower in the average token graphs compared to the normalized Shannon entropy graph. Generally, keyword subsequences naturally prioritize words carrying the most information, as reflected by the 10\% keyword category achieving the highest information density per percentage of full text across all categories. A consistent trend across all datasets is that, at the default 10\% threshold, NER words yield the lowest scores (apart from the title), followed by POS words. Furthermore, when selecting only 25\% of the keywords from the full text, the information density, measured by normalized Shannon entropy, is reduced by nearly half. Despite this reduction, we still achieve a comparable level of accuracy.

\end{document}